\magnification=1200

\centerline{\bf Zeros of some bi-orthogonal polynomials. }
\vskip 5mm

\centerline{Madan Lal Mehta\footnote*{E-mail: mehta@spht.saclay.cea.fr}}
\smallskip

\centerline{C.E. de Saclay, F-91191 Gif-sur-Yvette Cedex, France}
\vskip 1cm

{\bf Abstract.} Ercolani and McLaughlin have recently shown that
the zeros of the bi-orthogonal polynomials with the weight
$w(x,y)=\exp[-(V_1(x)+V_2(y)+2cxy)/2]$, relevant to a model of
two coupled hermitian matrices, are real and simple. We show that
their argument applies to the more general case of the weight
$(w_1*w_2*...*w_j)(x,y)$, a convolution of several weights of the
same form. This general case is relevant to a model of several
hermitian matrices coupled in a chain. Their argument also works
for the weight $W(x,y)=e^{-x-y}/(x+y)$, $0\le x,y<\infty$, and for a
convolution of several such weights. 
\vskip 1cm

{\bf 1. Introduction.} For a weight function $w(x,y)$ such that all 
the moments 
$$ M_{i,j} := \int w(x,y) x^i y^j dx dy \eqno (1.1)$$
exist and 
$$D_n := \det[M_{i,j}]_{i,j=0,1,...,n} \ne 0 \eqno (1.2) $$
for all $n\ge 0$, unique monic polynomials $p_n(x)$ and $q_n(x)$ 
of degree $n$ exist satisfying the bi-orthogonality relations
(a polynomial is called monic when the coefficient of
the highest degree is one)
$$\int w(x,y) p_n(x) q_m(y) dx dy = h_n \delta_{mn}. \eqno (1.3) $$
Just like the orthogonal polynomials
they can be expressed as determinants,  e.g. 
$$p_n(x)={1\over D_{n-1}}\det\left[ \matrix{
M_{0,0}&...&M_{0,n-1}&1\cr
M_{1,0}&...&M_{1,n-1}&x\cr
\vdots&\vdots&\vdots&\vdots\cr
M_{n,0}&...&M_{n,n-1}&x^n\cr   }\right] \eqno (1.4)$$
and have integral representations, e.g. 
$$ p_n(x) \propto \int \Delta_n({\bf x})\Delta_n({\bf y})
\prod_{j=1}^n (x-x_j) w(x_j,y_j) dx_j dy_j \eqno (1.5)$$
$$\Delta_n({\bf x}) := \prod_{1\le i<j\le n}(x_j-x_i), \hskip 5mm 
\Delta_n({\bf y}) := \prod_{1\le i<j\le n}(y_j-y_i). \eqno (1.6)$$

From limited numerical evidence for the weights
\item{(i)} $w(x,y)=\sin(\pi xy)$,\ \ \  $0\le x,y\le 1 $; 

\item{(ii)} $w(x,y)=\vert x-y\vert$,\ \ \ $-1\le x,y\le 1$; 

\item{(iii)} $w(x,y)=[1/(x+y)]\exp[-x-y]$,\ \ \ $0\le x,y<\infty$; 

\item{(iv)} $w(x,y)=\exp(-x^2-y^2-cxy)$,\ \ \ $-\infty<x,y<\infty$,
\ \ \ $0<c<2$; 

\noindent one might think that the zeros of the bi-orthogonal polynomials 
are real, simple, lie respectively in the $x$ or $y$-support of $w(x,y)$, 
interlace for successive $n$, ... 
\smallskip

Alas, this is not true in general as seen by the following example due to 
P. Deligne. If one takes 
$$ \eqalignno{ w(x,y) & = u(x,y) + v(x,y), & (1.7) \cr 
u(x,y) & = \left\{ \matrix{\delta(x-y), \ \ \ -1\le x,y\le 1, \cr
                 0, \ \ \ {\rm otherwise}, \cr} \right. & (1.8) \cr
v(x,y) & = {1\over 8} [\delta(x-1)\delta(y+2)+\delta(x+1)\delta(y-2)].
 & (1.9) \cr}$$
Then the zeros of $p_3(x)$ and $q_3(x)$ are complex. 
\smallskip

However, N.M. Ercolani and K.T.-R. Mclaughlin have recently [1] shown that 
with the weight function 
$$w_1(x,y)=\exp\left[-{1\over 2}V_1(x)-{1\over 2}V_2(y)-c_1xy\right] 
\eqno(1.10)$$ 
$(-\infty<x,y<\infty)$, $V_1$ and $V_2$ polynomials of positive
even degree, $c$
a small non-zero real constant, all the zeros of the bi-orthogonal 
polynomials $p_n(x)$ and $q_n(x)$ are real and simple. 
\smallskip

In this brief note we will show that their argument works for the following 
general case encountered for random hermitian matrices coupled in a linear
chain. Let
$V_j(x)$, $1\le j\le p$, be polynomials of positive even degree and $c_j$,
$1\le j<p$, be small real constants, none of them being zero  
(``small" so that all the moments $M_{i,j}$ defined below, eq.(1.13), 
exist.) Further let 
$$w_k(x,y):=\exp\left[-{1\over 2}V_k(x)-{1\over 2}V_{k+1}(y)-c_k xy\right] 
\eqno(1.11)$$ 
$$(w_{i_1}*w_{i_2}*...*w_{i_k})(\xi_1,\xi_{k+1}) :=\int w_{i_1}(\xi_1,\xi_2)
w_{i_2}(\xi_2,\xi_3)...w_{i_k}(\xi_k,\xi_{k+1})d\xi_2...d\xi_k
\eqno(1.12)$$
Moreover, assume that for all $i,j\ge 0$ 
$$M_{i,j}:=\int x^i(w_1*w_2*...*w_{p-1})(x,\,y)y^jdx dy \eqno (1.13)$$
exist.

\noindent{\bf Theorem.} Then monic polynomials $p_j(x)$ and $q_j(x)$ can
be uniquely defined by
$$\int p_j(x)(w_1*w_2*...*w_{p-1})(x,y)q_k(y)dx dy = h_j\delta_{jk} 
\eqno (1.14)$$
and all the zeros of $p_j(x)$ and of $q_j(x)$ are real and simple. 
\smallskip

The same argument works for any weight $W(x,y)$ such that 
$\det[W(x_i,y_j)]_{i,j=1,...,n}>0$ for $x_1<x_2<...<x_n$, 
$y_1<y_2<...<y_n$ and moments $M_{i,j}=\int W(x,y) x^i y^j dx dy$ 
exist for all $i,j\ge 0$. For example, if $W(x,y)=[1/(x+y)]\exp[-x-y]$, 
$0\le x,y<\infty$, then monic polynomials 
$p_j(x)$ can be uniquely defined by 
$$\int_0^\infty p_j(x)W(x,y)p_k(y)dx dy = h_j\delta_{jk} \eqno (1.15)$$
(here $W(x,y)$ is symmetric in $x$ and $y$ so that $p_j(x)=q_j(x)$)
and all the zeros of $p_j(x)$ are real, simple and non-negative.
\bigskip

{\bf 2. Results and proofs.}\ Here we essentially follow section 3 of 
reference [1].
With any monic polynomials $p_j(x)$ and $q_j(x)$ of degree $j$, let us 
write 
$$ \eqalignno{ P_{1,j}(x) & := p_j(x) & (2.1) \cr
P_{i,j}(x) & := \int p_j(\xi)(w_1*w_2*...*w_{i-1})(\xi,x) d\xi & \cr 
& := \int p_j(\xi) U_{Li}(\xi,x) d\xi, \hskip 5mm 1<i\le p & (2.2) \cr 
Q_{p,j}(x) & := q_j(x) & (2.3) \cr
Q_{i,j}(x) & := \int (w_i*w_{i+1}*...*w_{p-1})(x,\xi)q_j(\xi) d\xi & \cr 
& := \int U_{Ri}(x,\xi) q_j(\xi) d\xi \hskip 5mm 1\le i<p & (2.4) \cr}$$ 
\smallskip

{\bf Lemma 1.}\ For $x_1<x_2<...<x_n$, \ $y_1<y_2<...<y_n$, 
$$ \det\left[w_i(x_j,y_k)\right]_{j,k=1,...,n} >0. \eqno (2.5) $$ 
This is essentially eq. (40) of reference [1]. 
This can also be seen as follows. Let $X=[x_i\delta_{ij}]$ and 
$Y=[y_i\delta_{ij}]$ be two $n\times n$ diagonal matrices with diagonal 
elements $x_1$, ..., $x_n$ and $y_1$, ..., $y_n$ respectively. Then the 
integral of $\exp[-c\, {\rm tr}\,UXU^{-1}Y]$ over the $n\times n$ unitary 
matrices $U$ is given by [2] 
$$K{\det\left[\exp(-c\,x_iy_j)\right]_{i,j=1,...,n}
\over \Delta_n({\bf x})\Delta_n({\bf y})} \eqno (2.6)$$
where $K$ is a positive constant depending on $c$ and $n$. Hence 
$$
\exp\left[-{1\over 2}\sum_{j=1}^n \left(V_i(x_j)+V_{i+1}(y_j)\right)\right]
\int dU e^{-c_i{\rm tr}\,UXU^{-1}Y} 
=K{\det\left[w_i(x_j,y_k)\right]_{j,k=1,...,n}\over \Delta_n({\bf x})
\Delta_n({\bf y})} \eqno (2.7) $$
The left hand side is evidently positive while on the right hand side 
the denominator is positive since $x_1<x_2<...<x_n$ and $y_1<y_2<...<y_n$.
From this eq. (2.5) follows.
\smallskip

{\bf Lemma 2.}\ For $x_1<x_2<...<x_n$, \ \ \  $y_1<y_2<...<y_n$, 
$$\det\left[(w_{i_1}*w_{i_2}*...*w_{i_\ell})(x_j,y_k)
\right]_{j,k=1,...,n} >0 \eqno (2.8) $$
\smallskip

{\bf Proof.}\ Binet-Cauchy formula tells us that [3] 
$$\det\left[(w_{i_1}*w_{i_2})(x_j,y_k)\right]_{j,k=1,...,n}$$
is equal to 
$$\int_{\xi1<\xi_2<...<\xi_n}  
\det\left[w_{i_1}(x_j,\xi_k)\right]_{j,k=1,...,n}
.\det\left[w_{i_2}(\xi_j,y_k)\right]_{j,k=1,...,n} d\xi_1...d\xi_n 
\eqno (2.9) $$
By lemma 1 the integrand is every where positive, so lemma 2 is proved 
for the case $\ell=2$. The proof is now completed by induction on $\ell$, 
using again the Binet-Cauchy formula.
\smallskip

{\bf Lemma 3.}\ For any monic polynomial $p_j(x)$ of degree $j$, 
$P_{i,j}(x)$, $1\le i\le p$, has at most $j$ distinct real zeros. 
Similarly, for any monic polynomial $q_j(x)$ of degree $j$, $Q_{i,j}(x)$, 
$1\le i\le p$, has at most $j$ distinct real zeros. 
\smallskip

{\bf Proof.}\ Let, if possible, $z_1<z_2<...<z_m$, $m>j$, be the distinct 
real zeros of $P_{i,j}(x)$. Since 
$$P_{i,j}(x) = \sum_{k=0}^j a_k T_{i,k}(x), \eqno (2.10) $$
with 
$$ T_{i,k}(x) := \int \xi^k U_{Li}(\xi,x)d\xi,   \eqno (2.11) $$
$$P_{i,j}(z_\ell) = 0, \hskip 5mm \ell=1,2,...,m, 
\hskip 5mm m>j \eqno (2.12) $$ 
implies that 
$$ \eqalignno{ 0 & = \det \left[\matrix{ 
T_{i,0}(z_1) & T_{i,1}(z_1) & ... & T_{i,j}(z_1) \cr 
...          &...           & ... &...           \cr
T_{i,0}(z_{j+1}) & T_{i,1}(z_{j+1}) & ... & T_{i,j}(z_{j+1}) \cr }\right] 
& \cr
& = \int\det \left[ \matrix{
U_{Li}(\xi_1,z_1) & \xi_2U_{Li}(\xi_2,z_1) & ... & \xi_{j+1}^jU_{Li}(\xi_{j+1},z_1) 
\cr 
...          &...           & ... &...           \cr
U_{Li}(\xi_1,z_{j+1}) & \xi_2U_{Li}(\xi_2,z_{j+1}) & ... & \xi_{j+1}^j 
U_{Li}(\xi_{j+1},z_{j+1}) \cr }\right] d\xi_1 ... d\xi_{j+1} & \cr 
& = \int \xi_2\xi_3^2...\xi_{j+1}^j \det \left[ U_{Li}(\xi_k,z_\ell)
\right]_{k,\ell=1,...,j+1} d\xi_1 ... d\xi_{j+1} & (2.13) \cr}$$
or
$$ \int \det \left[U_{Li}(\xi_k,z_\ell)\right]_{k,\ell=1,...,j+1}
.\det\left[\xi_k^{\ell-1}\right]_{k,\ell=1,...,j+1}
d\xi_1...d\xi_{j+1} = 0 \eqno (2.14) $$
in contradiction to lemma 2. Thus $m$ can not be greater than $j$.

The proof for $Q_{i,j}(x)$ is similar.
\smallskip

{\bf Lemma 4.}\ Let the real constants $c_1$, ..., $c_{p-1}$, none of 
them being zero, be such that 
$$ M_{i,j} := \int x^i U_{Lp}(x,y) y^j dx dy 
\equiv \int x^i(w_1*w_2*...*w_{p-1})(x,y)y^j dx dy. \eqno (2.15) $$ 
exist for all $i,j\ge 0$. Then
$$ D_n := \det[M_{i,j}]_{i,j=1,...,n} \ne 0 \eqno (2.16) $$
for any $n\ge 0$.
\smallskip

{\bf Proof.}\ Let, if possible, $D_n=0$ for some $n$. Then 
$\sum_{j=0}^n M_{i,j}q_j = 0$, $q_j$ not all zero, and 
$$ \int x^i U_{Lp}(x,y)\sum_{j=0}^n q_j y^j dx dy = 0, \hskip 5mm 
i=0,\ 1, ..., n \eqno (2.17) $$ 
or
$$\int p_i(x) U_{Lp}(x,y) \sum_{j=0}^n q_j y^j dx dy = 0 \eqno (2.18) $$
for any polynomial $p_i(x)$ of degree $i\le n$. But
$$ \int U_{Lp}(x,y) \sum_{j=0}^n q_j y^j dy \eqno (2.19) $$
has at most $n$ distinct real zeros (lemma 3). So one can choose $p_i(x)$ 
such that 
$$p_i(x)\int U_{Lp}(x,y) \sum_{j=0}^n q_j y^j dy > 0 \eqno (2.20)$$ 
in contradiction to eq. (2.18).
So $D_n \ne 0$ and bi-orthogonal polynomials $p_j(x)$, $q_j(x)$ exist, see 
eqs. (1.4), (1.5).
\smallskip

{\bf Lemma 5.}\ Let $p_j(x)$, $q_j(x)$ be the bi-orthogonal polynomials,
eq. (1.14); or with the definitions (2.1)-(2.4) 
$$\int P_{i,j}(x) Q_{i,k}(x)dx = h_j\delta_{jk}, \hskip 5mm 1\le i\le p 
\eqno (2.21)$$
Then $P_{i,j}(x)$ has at least $j$ real distinct zeros of odd multiplicity. 
So does have $Q_{i,j}(x)$.
\smallskip

{\bf Proof.}\ Let, if possible, $z_1<z_2<...<z_m$, $m<j$, be the only real 
zeros of $P_{i,j}(x)$ of odd multiplicity. Set 
$$\eqalignno{ R(x) & = \det\left[ \matrix{ 
Q_{i,0}(x)&Q_{i,1}(x)&...&Q_{i,m}(x) \cr
Q_{i,0}(z_1)&Q_{i,1}(z_1)&...&Q_{i,m}(z_1) \cr
\vdots & \vdots & & \vdots \cr
Q_{i,0}(z_m)&Q_{i,1}(z_m)&...&Q_{i,m}(z_m) \cr} \right] & (2.22) \cr 
& = \int U_{Ri}(x,\xi)\sum_{k=0}^m \alpha_k \xi^k d\xi & (2.23) \cr}$$
with some constants $\alpha_k$ depending on $z_1$, ..., $z_m$. 

Since $m<j$, the bi-orthogonality gives 
$$\int P_{i,j}(x) R(x) dx =0. \eqno (2.24)$$
However, $R(x)$ can also be written as 
$$ \eqalignno{ R(x) 
& = \int \det \left[ \matrix{ 
U_{Ri}(x,\xi_0) & U_{Ri}(x,\xi_1)\xi_1 &... & U_{Ri}(x,\xi_m)\xi_m^m \cr
U_{Ri}(z_1,\xi_0)&U_{Ri}(z_1,\xi_1)\xi_1&...&U_{Ri}(z_1,\xi_m)\xi_m^m \cr
\vdots & \vdots & & \cr
U_{Ri}(z_m,\xi_0)&U_{Ri}(z_m,\xi_1)\xi_1&...&U_{Ri}(z_m,\xi_m)\xi_m^m \cr
} \right] d\xi_0 d\xi_1 ... d\xi_m   & \cr 
& = \int \det \left[ \matrix{ 
U_{Ri}(x,\xi_0) & U_{Ri}(x,\xi_1) &... & U_{Ri}(x,\xi_m) \cr
U_{Ri}(z_1,\xi_0)&U_{Ri}(z_1,\xi_1)&...&U_{Ri}(z_1,\xi_m) \cr
\vdots & \vdots & & \cr
U_{Ri}(z_m,\xi_0)&U_{Ri}(z_m,\xi_1)&...&U_{Ri}(z_m,\xi_m) \cr
} \right] \xi_1 \xi_2^2 ...\xi_m^m d\xi_0 d\xi_1 ... d\xi_m   & \cr 
& = {1\over (m+1)!} \int \det \left[ \matrix{ 
U_{Ri}(x,\xi_0) & U_{Ri}(x,\xi_1) &... & U_{Ri}(x,\xi_m) \cr
U_{Ri}(z_1,\xi_0)&U_{Ri}(z_1,\xi_1)&...&U_{Ri}(z_1,\xi_m) \cr
\vdots & \vdots & & \cr
U_{Ri}(z_m,\xi_0)&U_{Ri}(z_m,\xi_1)&...&U_{Ri}(z_m,\xi_m) \cr
} \right] \prod_{0\le r<s\le m}(\xi_s-\xi_r) d\xi_0 d\xi_1 ... d\xi_m   & \cr 
& = \int_{\xi_0<\xi_1<...<\xi_m} \det \left[ \matrix{ 
U_{Ri}(x,\xi_0) & U_{Ri}(x,\xi_1) &... & U_{Ri}(x,\xi_m) \cr
U_{Ri}(z_1,\xi_0)&U_{Ri}(z_1,\xi_1)&...&U_{Ri}(z_1,\xi_m) \cr
\vdots & \vdots & & \cr
U_{Ri}(z_m,\xi_0)&U_{Ri}(z_m,\xi_1)&...&U_{Ri}(z_m,\xi_m) \cr
} \right] \prod_{0\le r<s\le m}(\xi_s-\xi_r) d\xi_0 d\xi_1 ... d\xi_m   & 
\cr & & (2.25) \cr }$$
Thus $R(x)$ is represented as an integral whose integrand has a fixed 
sign determined by the relative ordering of the numbers $x$, 
$z_1$, $z_2$, ..., $z_m$ (lemma 2). It thus follows that $R(x)$ changes sign 
when $x$ passes through any of the points $z_k$, $k=1$, ..., $m$ 
and at no other value of $x$. In other words, $z_1$, ..., $z_m$ 
are the only real zeros of $R(x)$ having an odd multiplicity. 
And therefore $P_{i,j}(x)R(x)$ has a constant sign, so that 
$$\int P_{i,j}(x)R(x) dx \ne 0 \eqno (2.26)$$ 
in contradiction to (2.24). 
\smallskip

The proof for $Q_{i,j}(x)$ is similar.
\smallskip

As a consequence we have the integral representations of $P_{i,j}(x)$ for
$i>1$ and of $Q_{i,j}(x)$ for $i<p$ involving their respective zeros 
$$ \eqalignno{ P_{i,j}(x) & \propto \int \det \left[ \matrix{ 
U_{Li}(\xi_0,x)  & U_{Li}(\xi_1,x) &... & U_{Li}(\xi_j,x) \cr
U_{Li}(\xi_0,z_1)&U_{Li}(\xi_1,z_1)&... &U_{Li}(\xi_j,z_1) \cr
\vdots & \vdots & & \vdots \cr
U_{Li}(\xi_0,z_j)&U_{Li}(\xi_1,z_j)&...&U_{Li}(\xi_j,z_j) \cr
} \right] \prod_{0\le r<s\le j}(\xi_s-\xi_r) d\xi_0 d\xi_1 ... d\xi_j   & 
\cr & & (2.27) \cr 
Q_{i,j}(x) & \propto \int \det \left[ \matrix{ 
U_{Ri}(x,\xi_0) & U_{Ri}(x,\xi_1) &... & U_{Ri}(x,\xi_j) \cr
U_{Ri}(z_1,\xi_0)&U_{Ri}(z_1,\xi_1)&...&U_{Ri}(z_1,\xi_j) \cr
\vdots & \vdots & & \vdots \cr
U_{Ri}(z_j,\xi_0)&U_{Ri}(z_j,\xi_1)&...&U_{Ri}(z_j,\xi_j) \cr
} \right] \prod_{0\le r<s\le j}(\xi_s-\xi_r) d\xi_0 d\xi_1 ... d\xi_j   & 
\cr & & (2.28) \cr }$$
\smallskip
                     
Lemmas 3 and 5 tell us that if $p_j(x)$ and $q_j(x)$ are bi-orthogonal 
polynomials satisfying eq. (1.14), then $P_{i,j}(x)$ and $Q_{i,j}(x)$ 
each have exactly $j$ distinct real zeros of odd multiplicity. 
In particular, the zeros of the bi-orthogonal 
polynomials $p_j(x)\equiv P_{1,j}(x)$ and $q_j(x)\equiv Q_{p,j}(x)$ 
are real and simple. 
\smallskip

With a little more effort one can perhaps show that all the real zeros of 
$P_{i,j}(x)$ and of $Q_{i,j}(x)$ are simple. Other zeros, if any,  
must be complex. Whether the zeros of $p_j(x)$ ($q_j(x)$) interlace 
for successive $j$, remains an open question.
\bigskip 

{\bf 3. Bi-orthogonal polynomials with another weight.} 
\smallskip

For the weight $W(x,y)=[1/(x+y)]\exp[-x-y]$, $0\le x,y<\infty$, one 
can say as follows.  
\smallskip

{\bf Lemma 1'.} One has [4] 
$$\det[W(x_j,y_k)]_{j,k=1,...,n}=\exp\left[-\sum_{j=1}^n (x_j+y_j)\right] 
\Delta_n({\bf x}) \Delta_n({\bf y}) \prod_{j,k=1}^n (x_j+y_k)^{-1} \eqno (3.1)$$ 
which is evidently positive for $0\le x_1<x_2<...<x_n$, 
$0\le y_1<y_2<...<y_n$. 
\smallskip

{\bf Lemma 3'.} For any monic polynomial $p_j(x)$ of degree 
$j$, $P_j(x):=\int_0^\infty W(x,y) p_j(y) dy$  has at most $j$ distinct 
real non-negative zeros. 
\smallskip

In the proof of lemma 3, replace eqs. (2.10)-(2.14) by
$$P_j(x)=\sum_{k=0}^j a_k T_k(x), \eqno (3.2)$$ 
$$T_k(x)=\int_0^\infty \xi^k W(\xi,x) d\xi \eqno (3.3)$$
$$P_j(z_\ell)=0, \ \ \ \ell=1,2,...,m, \ \ \ m>j \eqno (3.4)$$
$$ \eqalignno{ 0 & = \det \left[\matrix{ 
T_0(z_1) & T_1(z_1) & ... & T_j(z_1) \cr 
...          &...           & ... &...           \cr
T_0(z_{j+1}) & T_1(z_{j+1}) & ... & T_j(z_{j+1}) \cr }\right] 
& \cr
& = \int_0^\infty \det \left[ \matrix{
W(\xi_1,z_1) & \xi_2 W(\xi_2,z_1) & ... & \xi_{j+1}^j W(\xi_{j+1},z_1) 
\cr 
...          &...           & ... &...           \cr
W(\xi_1,z_{j+1}) & \xi_2 W(\xi_2,z_{j+1}) & ... & \xi_{j+1}^j 
W(\xi_{j+1},z_{j+1}) \cr }\right] d\xi_1 ... d\xi_{j+1} & \cr 
& = \int \xi_2\xi_3^2...\xi_{j+1}^j \det \left[W(\xi_k,z_\ell)
\right]_{k,\ell=1,...,j+1} d\xi_1 ... d\xi_{j+1} & (3.5) \cr}$$
or
$$ \int \det \left[W(\xi_k,z_\ell)\right]_{k,\ell=1,...,j+1}
.\det\left[\xi_k^{\ell-1}\right]_{k,\ell=1,...,j+1}
d\xi_1...d\xi_{j+1} = 0 \eqno (3.6) $$
in contradiction to lemma 2'. Thus $m$ can not be greater than $j$.
\smallskip

{\bf Lemma 4'.} With 
$$M_{i,j}:=\int_0^\infty x^i W(x,y) y^j dx dy, \eqno (3.7)$$

$$D_n:=\det[M_{i,j}]_{i,j=0,1,...,n} \ne 0 \eqno (3.8)$$
for any $n\ge 0$.
\smallskip 

In the proof of lemma 4 replace everywhere $\int U_{Lp}(x,y)... $ by 
$\int_0^\infty W(x,y)...$. 
\smallskip 

{\bf Lemma 5'.} Let $p_j(x)$ be the (bi-orthogonal) polynomials satisfying 
$$\int_0^\infty W(x,y)p_j(x)p_k(y) dx dy =h_j\delta_{jk}. \eqno (3.9)$$ 
Then $P_j(x):=\int_0^\infty W(x,y)p_j(y) dy$ and $p_j(x)$ each have 
at least $j$ distinct real non-negative zeros of odd multiplicity. 
\smallskip 

Let, if possible, $0\le z_1<z_2<...<z_m$ $m<j$, be the only real 
non-negative zeros of $P_j(x)$ of odd multiplicity. Set 
$R(x)=\prod_{j=1}^m (x-z_j)$. Then as $m<j$, one has 
$$\int_0^\infty P_j(x)R(x) dx =0 \eqno (3.10)$$
But $P_j(x)$ and $R(x)$ change sign simultaneously as $x$ passes through 
the values $z_1$, ..., $z_m$ and at no other real positive value. So 
the product $P_j(x)R(x)$ never changes sign, in contradiction to (3.10). 
Therefore $P_j(x)$ has at least $j$ distinct real non-negative zeros of 
odd multiplicity. 
\smallskip 

To prove that $p_j(x)$ has at least $j$ distinct real non-negative zeros 
let if possible, $0\le z_1<z_2<...<z_m$, $m<j$, be the only such zeros. 
Set 
$$\eqalignno{ R(x) & = \det\left[ \matrix{ 
P_0(x)&P_1(x)&...&P_m(x) \cr
P_0(z_1)&P_1(z_1)&...&P_m(z_1) \cr
\vdots & \vdots & & \vdots \cr
P_0(z_m)&P_1(z_m)&...&P_m(z_m) \cr} \right] & \cr 
& = \int_0^\infty W(x,\xi)\sum_{k=0}^m \alpha_k \xi^k d\xi & (3.11) \cr}$$
with some constants $\alpha_k$ depending on $z_1$, ..., $z_m$. 

Since $m<j$, the bi-orthogonality gives 
$$\int_0^\infty P_j(x) R(x) dx =0. \eqno (3.12)$$
But  
$$ \eqalignno{ R(x) 
& \propto \int_0^\infty \det \left[ \matrix{ 
W(x,\xi_0) & W(x,\xi_1) &... & W(x,\xi_m) \cr
W(z_1,\xi_0)&W(z_1,\xi_1)&...&W(z_1,\xi_m) \cr
\vdots & \vdots & & \cr
W(z_m,\xi_0)&W(z_m,\xi_1)&...&W(z_m,\xi_m) \cr
} \right] \prod_{0\le r<s\le m}(\xi_s-\xi_r) d\xi_0 d\xi_1 ... d\xi_m  & \cr  
& & (3.13) \cr}$$
which says that $z_1$, ..., $z_m$ are the only distinct real non-negative 
zeros of $R(x)$ and therefore $p_j(x)R(x)$ has a constant sign, in 
contradiction to (3.12). 
\bigskip

{\bf Conclusion.} We have shown with the arguments of Ercolani and
McLaughlin that                                                                      
if the weight $w(x,y)$ is such that $\det[w(x_i,y_j)]_{i,j=1,...,n}>0$
for $x_1<x_2<...<x_n$, $y_1<y_2<...<y_n$ and moments
$\int w(x,y)x^i y^j dx dy$ exist  for all $i,j\ge 0$, then
bi-orthogonal polynomials exist and their zeros are real, simple and lie
in the respective supports of the  weight $w(x,y)$.
The same is true for a weight which is a convolution of several such weights.
\bigskip

{\bf Acknowlegements.} I am thankful to P.M. Bleher to supply me a copy
of the preprint [1].
\bigskip

{\bf References.} 
\smallskip

\item{[1]} N.M. Ercolani and K.T.-R. Mclaughlin, Asymptotic and integrable 
structures for bi-orthogonal polynomials associated to a random two 
matrix model (preprint, to appear)

\item{[2]} See e.g. M.L. Mehta, {\it Random matrices}, Academic Press, 
new York (1991) Appendix A.5

\item{[3]} See e.g. M.L. Mehta, {\it Matrix theory}, Les Editions de 
Physique, Z.I. de Courtaboeuf, 91944 Les Ulis Cedex, France, (1989) \S 3.7

\item{[4]} See e.g. reference 3, \S 7.1.3 

\end